\documentclass{llncs}

\usepackage{xspace}
\usepackage{fancybox}
\usepackage[utf8]{inputenc}
\usepackage{url}
\usepackage{booktabs}
\usepackage{multirow}
\usepackage{multicol}

\usepackage{microtype}
\usepackage{xcolor}
\usepackage{enumitem,amsmath,amssymb}
\usepackage{breakurl}    % used for \url and \burl
\usepackage[linesnumbered,boxed,noline,noend]{algorithm2e}

\usepackage{tikz}
\usepackage{subfig}
\usepackage{array,booktabs,multirow}
\usepackage{placeins}

\usepackage{graphicx}
\usepackage{wrapfig}

\usepackage{logictools}
\usepackage{prooftheory}
\usepackage{comment}
\usepackage{mathenvironments}
\usepackage{drawproof}
\usepackage{proof}

\newcommand{\class}[1]{\texttt{#1}}

\title{ 
  Scavenger 0.1: \\
  A Theorem Prover Based on Conflict Resolution
}

\author{
  Daniyar Itegulov\inst{1}
  \and
  John Slaney\inst{2}
  \and 
  Bruno Woltzenlogel Paleo\inst{2}
  \thanks{Author order is alphabetical by surname.}
}

\authorrunning{D.\~Itegulov \and B.\~Woltzenlogel Paleo}

\institute{
  ITMO University, St. Petersburg, Russia \\
  \email{ditegulov@gmail.com}
  \and 
  Australian National University, Canberra, Australia \\
  \email{john.slaney@anu.edu.au} \\
  \email{bruno.wp@gmail.com}
}

\begin{document}

\maketitle

\begin{abstract}
This paper introduces \scavenger, the first theorem prover for pure first-order logic without equality based on the new conflict resolution calculus. Conflict resolution has a restricted resolution inference rule that resembles (a first-order generalization of) unit propagation as well as a rule for assuming decision literals and a rule for deriving new clauses by (a first-order generalization of) conflict-driven clause learning.
\end{abstract}

\setcounter{footnote}{0}

\section{Introduction}

The outstanding efficiency of current propositional \textsc{Sat}-solvers naturally raises the question 
of whether it would be possible to employ similar ideas for automating first-order logical reasoning. 
The recent \emph{Conflict Resolution} calculus\footnote{Not to be confused with the homonymous calculus for linear rational inequalities~\cite{KorovinTVConflictResolution}.} (\CR)~\cite{ConflictResolution} can be regarded as a crucial initial step to answer this question. From a proof-theoretical perspective, \CR generalizes (to first-order logic) the two main mechanisms on which modern \textsc{Sat}-solvers are based: unit propagation and conflict-driven clause learning. The calculus is sound and refutationally complete, and \CR derivations are isomorphic to implication graphs.

This paper goes one step further by defining proof search algorithms for \CR. Familiarity with the propositional CDCL procedure~\cite{CDCL} is assumed, even though it is briefly sketched in Section~\ref{sec:CDCL}. The main challenge in lifting this procedure to first-order logic is that, unlike in propositional logic, first-order unit propagation does not always terminate and true clauses do not necessarily have uniformly true literals (cf. Section~\ref{sec:challenges}). Our solutions to these challenges are discussed in Section~\ref{sec:search} and Section~\ref{sec:Implementation}, and experimental results are presented in Section~\ref{sec:experiments}.

\paragraph{Related Work:} 

\CR's unit-propagating resolution rule can be traced back to unit-resulting resolution \cite{URR}. Other attempts to lift DPLL~\cite{DavisPutnam,DLL} or CDCL~\cite{CDCL} to first-order logic include \emph{Model Evolution}~\cite{BaumgartnerFODPLL,BaumgartnerTinelli,Baumgartner,BaumgartnerTinelliLemmaLearning}, \emph{Geometric Resolution}~\cite{GeometricResolution}, \emph{Non-Redundant Clause Learning}~\cite{AlagiWeidenbach} and the \emph{Semantically-Guided Goal Sensitive procedure}~\cite{BonacinaPlaisted1,BonacinaPlaisted2,BonacinaPlaisted3,BonacinaPlaisted4}. A brief summary of these approaches and a comparison with \CR can be found in~\cite{ConflictResolution}. 
Furthermore, many architectures~\cite{Equinox,iProver,InstGen,AVATAR,Satallax} for first-order and higher-order theorem proving use a \textsc{Sat}-solver as a black box for propositional reasoning, without attempting to lift it; and \emph{Semantic Resolution}~\cite{SCOtt1,SCOtt2} is yet another related approach that uses externally built first-order models to guide resolution.

\section{Propositional CDCL}
\label{sec:CDCL}

During search in the propositional case, a \textsc{Sat}-solver keeps a model (a.k.a. trail) consisting of a (conjunctive) list of decision literals and propagated literals. Literals of unit clauses are automatically added to the trail, and whenever a clause has only one literal that is not falsified by the current model, this literal is added to the model (thereby satisfying that clause). This process is known as \emph{unit-propagation}. If unit propagation reaches a conflict (i.e. a situation where the dual of a literal already contained in the model would have to be added to it), the \textsc{Sat}-solver backtracks, removing from the model decision literals responsible for the conflict (as well as propagated literals entailed by the removed decision literals) and deriving, or learning, a conflict-driven clause consisting\footnote{In practice, optimizations (e.g. 1UIP) are used, and more sophisticated clauses, which are not just disjunctions of duals of the decision literals involved in the conflict, can be derived. But these optimizations are inessential to the focus of this paper.} of duals of the decision literals responsible for the conflict (or the empty clause, if there were no decision literals). If unit propagation terminates without reaching a conflict and all clauses are satisfied by the model, then the input clause set is satisfiable. If some clauses are still not satisfied, the \textsc{Sat}-solver chooses and assigns another decision literal, adding it to the trail, and satisfying the clauses that contain it.

\section{Conflict Resolution}

\begin{figure}
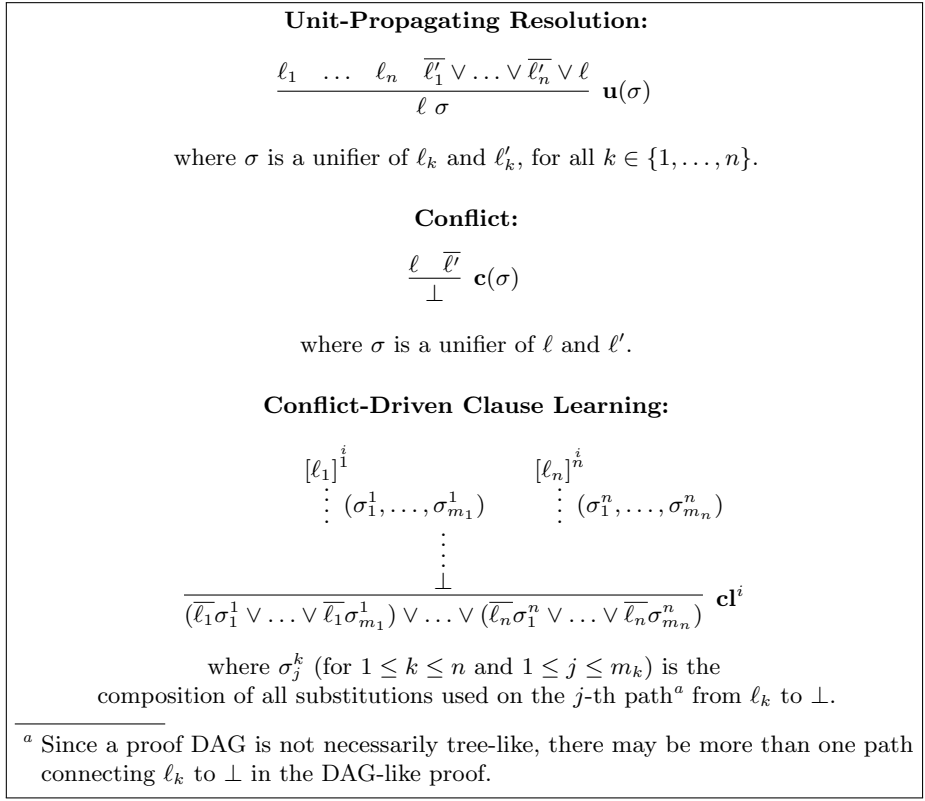

\begin{calculus}
\centering

\textbf{Unit-Propagating Resolution:}
$$
\infer[\upr{\sigma}]{\ell~\sigma}{\ell_1 & \ldots & \ell_n & \dual{\ell'_1} \vee \ldots \vee \dual{\ell'_n} \vee \ell}
$$

where $\sigma$ is a unifier of $\ell_k$ and $\ell'_k$, for all $k \in \{1, \ldots, n \}$.

\bigskip
%\bigskip

\textbf{Conflict:}
$$
\infer[\con{\sigma}]{\bot}{\ell & \dual{\ell'}}
$$

where $\sigma$ is a unifier of $\ell$ and $\ell'$.

%\bigskip
\bigskip

\textbf{Conflict-Driven Clause Learning:}
$$
\infer[\cdcl^{i}]
{(\dual{\ell_1} \sigma^1_1 \vee \ldots \vee \dual{\ell_1} \sigma^1_{m_1}) \vee \ldots \vee (\dual{\ell_n} \sigma^n_1 \vee \ldots \vee \dual{\ell_n} \sigma^n_{m_n})}
{\infer*{\bot}{\infer*[(\sigma_1^1,\ldots,\sigma_{m_1}^1)]{}{[\ell_1]^{\dls{i}{1} } } &  & \infer*[(\sigma_1^n,\ldots,\sigma_{m_n}^n)]{}{[\ell_n]^{\dls{i}{n} } } }}
$$

where $\sigma^k_j$ (for $1 \leq k \leq n$ and $1 \leq j \leq m_k$) is the \\
composition of all substitutions used on the $j$-th path\footnote{Since a proof DAG is not necessarily tree-like, there may be more than one path connecting $\ell_k$ to $\bot$ in the DAG-like proof.} from $\ell_k$ to $\bot$.

\end{calculus}
\caption{The Conflict Resolution Calculus \CR}
\label{fig:CR}
\end{figure}

The inference rules of the conflict resolution calculus \CR are shown in Figure~\ref{fig:CR}. The \emph{unit propagating resolution} rule is a chain of restricted resolutions with unit clauses as left premises and a unit clause as final conclusion. \emph{Decision literals} are denoted by square brackets, and the \emph{conflict-driven clause learning} rule infers a new clause consisting of negations of instances of decision literals used to reach a conflict (a.k.a. the empty clause $\bot$). A clause learning inference is said to discharge the decision literals that it uses. As in the resolution calculus, \CR derivations are directed acyclic graphs that are not necessarily tree-like. A \CR refutation is a \CR derivation of $\bot$ with no undischarged decision literals. 

From a natural deduction point of view, a unit propagating resolution rule can be regarded as a chain of implication eliminations taking unification into account, whereas decision literals and conflict driven clause learning are reminiscent of, respectively, assumptions and chains of negation introductions, also generalized to first-order through unification. Therefore, \CR can be considered a first-order hybrid of resolution and natural deduction.

\section{Lifting Challenges}
\label{sec:challenges}

First-order logic presents many new challenges for methods based on propagation and decisions, of which the following can be singled out:

\paragraph{\textbf{(1)} non-termination of unit-propagation:}
In first-order logic, unit propagation may never terminate. For example, the clause set $\{ p(a), \neg p(X) \lor p(f(X)), q \lor r, \neg q \lor r, q \lor \neg r, \neg q \lor \neg r \}$ is clearly unsatisfiable, because there is no assignment of $p
$ and $q$ to \emph{true} or \emph{false} that would satisfy all the last four clauses. However, unit propagation would derive the following infinite sequence of units, by successively resolving $\neg
p(X) \lor p(f(X))$ with previously derived units, starting with $p(a)$: $\{ p(f(a)), p(f(f(a))), \ldots, p(f(\ldots(f(a))\ldots)),\ldots\}$. Consequently, a proof search strategy that would wait for unit propagation to terminate before making decisions would never be able to conclude that the given clause set is unsatisfiable.

\paragraph{\textbf{(2)} absence of uniformly true literals in satisfied clauses:} 
While in the propositional case, a clause that is true in a model always has at least one literal that is true in that model, this is not so in first-order logic, because shared variables create dependencies between literals. For instance, the clause set $\{p(X) \vee q(X), \neg p(a), p(b), q(a), \neg q(b) \}$ is satisfiable, but there is no model where $p(X)$ is uniformly true (i.e. true for all instances of $X$) or $q(X)$ is uniformly true. 

\paragraph{\textbf{(3)} propagation without satisfaction:}
In the propositional case, when only one literal of a clause is not false in the model, this literal is propagated and added to the model, and the clause necessarily becomes true in the model and does not need to be considered in propagation anymore, at least until backtracking. In the first-order case, on the other hand, a clause such as $p(X) \vee q(X)$ would propagate the literal $q(a)$ in a model containing $\neg p(a)$, but $p(X) \vee q(X)$ does not become true in a model where $q(a)$ is true. It must remain available for further propagations. If, for instance, the literal $\neg p(b)$ is added to the model, the clause will be used again to propagate $q(b)$.

\paragraph{\textbf{(4)} quasi-falsification without propagation:}
A clause is \emph{quasi-falsified} by a model iff all but one of its literals are false in the model. 
In first-order logic, in contrast to propositional logic, it is not even the case that a clause will necessarily propagate a literal when only one of its literals is not false in the model. For instance, the clause $p(X) \vee q(X) \vee r(X)$ is quasi-falsified in a model containing $\neg p(a)$ and $\neg q(b)$, but no instance of $r(X)$ can be propagated. 

The first two challenges affect search in a conceptual level, and solutions are discussed in Section~\ref{sec:search}. The last two prevent a direct first-order generalization of the data structures (e.g. \emph{watched literals}) that make unit propagation so efficient in the propositional case. Partial solutions are discussed in Section~\ref{sec:Implementation}.

% \paragraph{\textbf{(5)} infinite models:}
% In first-order logic, some satisfiable clause sets may admit only infinite models. 
% %In such cases, a naive model construction based on unit propagation may never terminate. 
% %For instance, this would occur on the following clause set: $\{ p(a), \neg p(X) \vee p(f(X)) \}$ \marginpar{replace this by a clause set that has only infinite models in general, and not only when restricted to Herbrand models. E.g. this could be a clause set encoding injectivity and not surjectivity without using equality}. 
% Although it would be possible to check for certain conditions and detect some cases of satisfiable clause sets with infinite models, it would be hopeless to try to detect all such cases, since this would entail a decision procedure for unsatisfiability in first-order logic, which is known to be undecidable.

\section{First-Order Model Construction and Proof Search}
\label{sec:search}

Despite the fundamental differences between propositional and first-order logic described in the previous section, the first-order algorithms presented aim to adhere as much as possible to the propositional procedure sketched in the Section~\ref{sec:CDCL}. As in the propositional case, the model under construction is a (conjunctive) list of literals, but literals may now contain (universal) variables. If a literal $\ell[X]$ is in a model $M$, then any instance $\ell[t]$ is said to be true in $M$. Note that checking that a literal $\ell$ is true in a model $M$ is more expensive in first-order logic than in propositional logic: whereas in the latter it suffices to check that $\ell$ is in $M$, in the former it is necessary to find a literal $\ell'$ in $M$ and a substitution $\sigma$ such that $\ell = \ell' \sigma$. A literal $\ell$ is said to be \emph{strongly true} in a model $M$ iff $\ell$ is in $M$.

There is a straightforward solution for the second challenge (i.e. the absence of uniformly true literals in satisfied clauses): a clause is satisfied by a model $M$ iff all its relevant instances have a literal that is true in $M$, where an instance is said to be relevant if it substitutes the clause's variables by terms that occur in $M$. Thus, for instance, the clause $p(X) \vee q(X)$ is satisfied by the model $[\neg p(a), p(b), q(a), \neg q(b)]$, because both relevant instances $p(a) \vee q(a)$ and $p(b) \vee q(b)$ have literals that are true in the model.
However, this solution is costly, because it requires the generation of many instances. Fortunately, in many (though not all) cases, a satisfied clause will have a literal that is true in $M$, in which case the clause is said to be \emph{uniformly satisfied}. Uniform satisfaction is cheaper to check than satisfaction. However, a drawback of uniform satisfaction is that the model construction algorithm may repeatedly attempt to satisfy a clause that is not uniformly satisfied, by choosing one of its literals as a decision literal. For example, the clause $p(X) \vee q(X)$ is not uniformly satisfied by the model $[\neg p(a), p(b), q(a), \neg q(b)]$. Without knowing that this clause is already satisfied by the model, the procedure would try to choose either $p
(X)$ or $q(X)$ as decision literal. But both choices are \emph{useless decisions}, because they would lead to conflicts with conflict-driven clauses equal to a previously derived clause or to a unit clause containing a literal that is part of the current model. A clause is said to be \emph{weakly satisfied} by a model $M$ if and only if all its literals are useless decisions.

Because of the first challenge (i.e. the non-termination of unit-propagation in the general first-order case), it is crucial to make decisions \emph{during} unit propagation. In the example given in item 1 of Section~\ref{sec:challenges}, for instance, deciding $q$ at any moment would allow the propagation of $r$ and $\neg r$ (respectively due to the 4th and 6th clauses), triggering a conflict. The learned clause would be $\neg q$ and it would again trigger a conflict by the propagation of $r$ and $\neg r$ (this time due to the 3rd and 5th clauses). As this last conflict does not depend on any decision literal, the empty clause is derived and thus the clause set is refuted. The question is how to interleave decisions and propagations. One straightforward approach is to keep track of the \emph{propagation depth}\footnote{Because of the isomorphism between implication graphs and subderivations in Conflict Resolution~\cite{ConflictResolution}, the propagation depth is equal to the corresponding subderivation's \emph{height}, where initial axiom clauses and learned clauses have height $0$ and the height of the conclusion of a unit-propagating resolution inference is $k + 1$ where $k$ is the maximum height of its unit premises.} in the implication graph: any decision literal or literal propagated by a unit clause has propagation depth $0$; any other literal has propagation depth $k + 1$, where $k$ is the maximum propagation depth of its predecessors. Then propagation is performed exhaustively only up to a propagation depth threshold $h$. A decision literal is then chosen and the threshold is incremented. Such \emph{eager decisions} guarantee that a decision will eventually be made, even if there is an infinite propagation path. However, eager decisions may also lead to spurious conflicts generating useless conflict-driven clauses. For instance, the clause set $\{ 1: p(a), 2: \neg p(X) \vee p(f(X)), 3: \neg p(f(f(f(f(f(f(a))))))), 4: \neg r(X) \vee q(X), 5: \neg q(g(X)) \vee \neg p(X), 6: z(X) \vee r(X) \}$ (where clauses have been numbered for easier reference) is unsatisfiable, because a conflict with no decisions can be obtained by propagating $p(a)$ (by 1), and then $p(f(a))$, $p(f(f(a)))$, \ldots, $p(f(f(f(f(f(f(a)))))))$, (by 2, repeatedly), which conflicts with $\neg p(f(f(f(f(f(f(a)))))))$ (by 3). But the former propagation has depth $6$. If the propagation depth threshold is lower than $6$, a decision literal is chosen before that conflict is reached. If $r(X)$ is chosen, for example, in an attempt to satisfy the sixth clause, there are propagations (using $r(X)$ and clauses 1, 4, 5 and 6) with depth lower than the threshold and reaching a conflict that generates the clause $\neg r(g(a))$, which is useless for showing unsatisfiability of the whole clause set. This is not a serious issue, because useless clauses are often generated in conflicts with non-eager decisions as well. Nevertheless, this example suggests that the starting threshold and the strategy for increasing the threshold have to be chosen wisely, since the performance may be sensitive to this choice.

Interestingly, the problem of non-terminating propagation does not manifest in fragments of first-order logic where infinite unit propagation paths are impossible. A well-known and large fragment is the \emph{effectively propositional} (a.k.a. \emph{Bernays-Sch\"onfinkel}) class, consisting of sentences with prenex forms that have an $\exists^* \forall^*$ quantifier prefix and no function symbols. For this fragment, a simpler proof search strategy that only makes decisions when unit propagation terminates, as in the propositional case, suffices. Infinite unit propagation paths do not occur in the effectively propositional fragment because there are no function symbols and hence the term depth\footnote{The depth of constants and variables is zero and the depth of a complex term is $k+1$ when $k$ is the maximum depth of its proper subterms.} does not increase arbitrarily. Whenever the term depth is bounded, infinite unit propagation paths cannot occur, because there are only finitely many literals with bounded term depth (given the finite set of constant, function and predicate symbols with finite arity occurring in the clause set). 

The insight that term depth is important naturally suggests a different approach for the general first-order case: instead of limiting the propagation depth, limit the \emph{term depth} instead, allowing arbitrarily long propagations as long as the term depth of the propagated literals are smaller than the current term depth threshold. A literal is propagated only if its term depth is smaller than the threshold. New decisions are chosen when the term-depth-bounded propagation terminates and there are still clauses that are not uniformly satisfied. As before, eager decisions may lead to spurious conflicts, but bounding propagation by term depth seems intuitively more sensible than bounding it by propagation depth.

\section{Implementation Details}
\label{sec:Implementation}

\scavenger is implemented in Scala and its source code and usage instructions are available in \url{https://gitlab.com/aossie/Scavenger}. Its packrat combinator parsers are able to parse TPTP CNF files~\cite{TPTP}. Although \scavenger is a first-order prover, every logical expression is converted to a simply typed lambda expression, implemented by the abstract class \class{E} with concrete subclasses \class{Sym}, \class{App} and \class{Abs} for, respectively, \emph{symbols}, \emph{applications} and \emph{abstractions}. A trait \class{Var} is used to distinguish \emph{variables} from other symbols. Scala's \emph{case classes} are used to make \class{E} behave like an algebraic datatype with (pattern-matchable) constructors. The choice of simply typed lambda expressions is motivated by the intention to generalize \scavenger to multi-sorted first-order logic and higher-order logic and support TPTP TFF and THF in the future. Every clause is internally represented as an immutable two-sided sequent consisting of a set of positive literals (succedent) and a set of negative literals (antecedent).

When a problem is unsatisfiable, \scavenger can output a \CR refutation internally represented as a collection of \class{ProofNode} objects, which can be instances of the following immutable classes: \class{UnitPropagatingResolution}, \class{Conflict}, \class{ConflictDrivenClauseLearning}, \class{Axiom}, \class{Decision}. The first three classes correspond directly to the rules shown in Figure~\ref{fig:CR}. \class{Axiom} is used for leaf nodes containing input clauses, and \class{Decision} represents a fictive rule holding decision literals. Each class is responsible for checking, typically through \texttt{require} statements, the soundness conditions of its corresponding inference rule. The \class{Axiom}, \class{Decision} and \class{ConflictDrivenClauseLearning} classes are less than 5 lines of code each. \class{Conflict} and \class{UnitPropagatingResolution} are respectively 15 and 35 lines of code. The code for analyzing conflicts, traversing the subderivations (conflict graphs) and finding decisions that contributed to the conflict, is implemented in a superclass, and is 17 lines long.

The following three variants of \scavenger were implemented:
\begin{itemize}
\item EP-\scavenger: aiming at the effectively propositional fragment, propagation is not bounded, and decisions are made only when propagation terminates.

\item PD-\scavenger: Propagation is bounded by a propagation depth 
threshold starting at $0$. Input clauses are assigned depth $0$. Derived clauses and propagated literals obtained while the depth threshold is $k$ are assigned depth $k+1$. The threshold is incremented whenever every input clause that is neither uniformly satisfied nor weakly satisfied is used to derive a new clause or to propagate a new literal. If this is not the case, a decision literal is chosen (and assigned depth $k+1$) to uniformly satisfy one of the clauses that is neither uniformly satisfied nor weakly satisfied. 

\item TD-\scavenger: Propagation is bounded by a term depth 
threshold starting at $0$. When propagation terminates, a stochastic choice between either selecting a decision literal or incrementing the threshold is made with probability of 50\% for each option. 
Only uniform satisfaction of clauses is checked.
\end{itemize}

The third and fourth challenges discussed in Section~\ref{sec:challenges} are critical for performance, because they prevent a direct first-order generalization of data structures such as \emph{watched literals}, which enables efficient detection of clauses that are ready to propagate literals. Without knowing exactly which clauses are ready to propagate, \scavenger (in its three variants) loops through all clauses with the goal of using them for propagation. However, actually trying to use a given clause for propagation is costly. In order to avoid this cost, \scavenger performs two quicker tests. Firstly, it checks whether the clause is uniformly satisfied (by checking whether one of its literals belongs to the model). If it is, then the clause is dismissed. This is an imperfect test, however. Occasionally, some satisfied clauses will not be dismissed, because (in first-order logic) not all satisfied clauses are uniformly satisfied. Secondly, for every literal $\ell$ of every clause, \scavenger keeps a set of decision literals and propagated literals that are unifiable with $\ell$. A clause $c$ is quasi-falsified when at most one literal of $c$ has an empty set associated with it. This is a rough analogue of watched literals for detecting quasi-falsified clauses. Again, this is an imperfect test, because (in first-order logic) not all quasi-falsified clauses are ready to propagate. Despite the imperfections of these tests, they do reduce the number of clauses that need to be considered for propagation, and they are quick and simple to implement. 

Overall, the three variants of \scavenger listed above have been implemented concisely. Their main classes are only 168, 342 and 176 lines long, respectively, and no attempt has been made to increase efficiency at the expense of the code's readability and pedagogical value. Premature optimization would be inappropriate for a first proof-of-concept.

\scavenger still has no sophisticated backtracking and restarting mechanism, as propositional \textsc{Sat}-solvers do. When \scavenger reaches a conflict, it restarts almost completely: all derived conflict-driven clauses are kept, but the model under construction is reset to the empty model.

\section{Experiments}
\label{sec:experiments}

Experiments were conducted\footnote{Raw experimental data are available at \url{https://doi.org/10.5281/zenodo.293187}.} in the StarExec cluster~\cite{StarExec} to evaluate \scavenger's performance on TPTP v6.4.0 benchmarks in CNF form and without equality. 
%There were 1606 unsatisfiable benchmarks (of which 572 in the effectively propositional fragment) and TODO satisfiable benchmarks (of which 185 effectively propositional). 
For comparison, all other 21 provers available in StarExec's TPTP community and suitable for CNF problems without equality were evaluated as well. For each job pair, the timeouts were 300 CPU seconds and 600 Wallclock seconds.

%The 29 provers satisfying these criteria included: LEO-II, ZenonModulo, Geo-III, SOS, Otter, (2 variants of) Beagle, E-KRHyper, Zipperpin, Prover9, Metis, (3 variants of) Darwin, SNARK, Bliksem, PEPR, GrAnDe, CVC4, Paradox, ET, E, Z3, iProver and (5 variants of) Vampire. 

\newcommand{\s}[1]{\begin{small}#1\end{small}}

\begin{table}

\begin{multicols}{2}

%\begin{center}
\noindent
  \begin{tabular*}{0.485\textwidth}{  l @{\extracolsep{\fill}} r  r  }
    \toprule
                    &  \multicolumn{2}{c}{\textbf{Problems Solved}} \\
\textbf{Prover}     &  \emph{EPR}  & \emph{All} \\ \midrule
\s{PEPR-0.0ps} & 432 & 432 \\ 
\s{GrAnDe-1.1} & 447 & 447 \\ 
\s{Paradox-3.0} & 467 & 506 \\ 
\s{ZenonModulo-0.4.1} & 315 & 628 \\ 
\s{\emph{\textbf{TD-Scavenger}}} & \textbf{\emph{350}} & \textbf{\emph{695}} \\ 
\s{\emph{\textbf{PD-Scavenger}}} & \textbf{\emph{252}} & \textbf{\emph{782}} \\ 
\s{Geo-III-2016C} & 344 & 840 \\ 
\s{\emph{\textbf{EP-Scavenger}}} & \textbf{\emph{349}} & \textbf{\emph{891}} \\ 
\s{Metis-2.3} & 404 & 950 \\ 
\s{Z3-4.4.1} & 507 & 1027 \\ 
\s{Zipperpin-FOF-0.4}  & 400 & 1029 \\ 
\s{Otter-3.3}  & 362 & 1068 \\
    \bottomrule
  \end{tabular*}
%\end{center}
%

%\begin{table}
%\begin{center}
\noindent
  \begin{tabular*}{0.485\textwidth}{  l @{\extracolsep{\fill}} r  r  }
    \toprule
                    &  \multicolumn{2}{c}{\textbf{Problems Solved}} \\
\textbf{Prover}     &  \emph{EPR}  & \emph{All} \\ \midrule
\s{Bliksem-1.12}  & 424 & 1107 \\  
\s{SOS-2.0}  & 351 & 1129 \\
\s{CVC4-FOF-1.5.1}  & 452 & 1145 \\ 
\s{SNARK-20120808}  & 417 & 1150 \\ 
\s{Beagle-0.9.47} & 402 & 1153 \\ 
\s{E-Darwin-1.5}  & 453 & 1213 \\ 
\s{Prover9-1109a}  & 403 & 1293 \\ 
\s{Darwin-1.4.5}  & 508 & 1357 \\ 
\s{iProver-2.5}  & 551 & 1437 \\ 
\s{ET-0.2}  & 486 & 1455 \\ 
\s{E-2.0}  & 489 & 1464 \\ 
\s{Vampire-4.1} & 540 & 1524 \\
    \bottomrule
  \end{tabular*}
%\end{center}
%\end{table}

\end{multicols}
\caption{Number of problems solved by each prover}
\label{table:Totals}
\end{table}

\begin{figure}
\includegraphics[width=\textwidth]{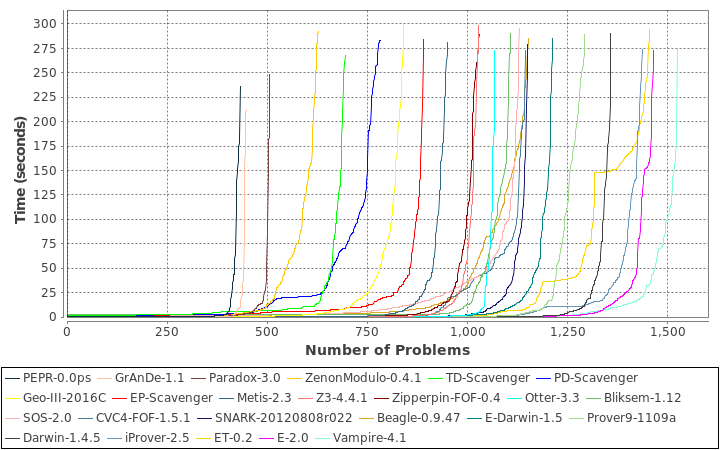}
\caption{Performance on all benchmarks (provers ordered by performance)}
\label{fig:Unsat}
\end{figure}

\begin{figure}
\includegraphics[width=\textwidth]{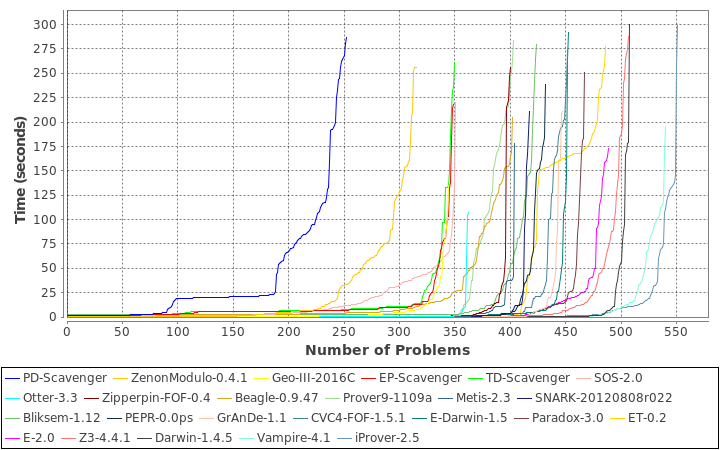}
\caption{Performance on EPR benchmarks only (provers ordered by performance)}
\label{fig:EPUnsat}
\end{figure}

Table \ref{table:Totals} shows how many of the 1606 unsatisfiable CNF problems and 572 effectively propositional (EPR) unsatisfiable CNF problems each theorem prover solved; and figures \ref{fig:Unsat} and \ref{fig:EPUnsat} shows the performance in more detail. For a first implementation, the best variants of \scavenger show an acceptable performance. All variants of \scavenger outperformed PEPR, GrAnDe, DarwinFM, Paradox and ZenonModulo; and EP-\scavenger additionally outperformed Geo-III. On the effectively propositional propblems, TD-\scavenger outperformed LEO-II, ZenonModulo and Geo-III, and solved only 1 problem less than SOS-2.0 and 12 less than Otter-3.3. Although Otter-3.3 has long ceased to be a state-of-the-art prover and has been replaced by Prover9, the fact that \scavenger solves almost as many problems as Otter-3.3 is encouraging, because Otter-3.3 is a mature prover with 15 years of development, implementing (in the C language) several refinements of proof search for resolution and paramodulation (e.g. orderings, set of support, splitting, demodulation, subsumption)~\cite{Otter,OtterManual}, whereas \scavenger is a yet unrefined and concise implementation (in Scala) of a comparatively straightforward search strategy for proofs in the Conflict Resolution calculus, completed in slightly more than 3 months. Conceptually, Geo-III (based on Geometric Resolution) and Darwin (based on Model Evolution) are the most similar to \scavenger. While \scavenger already outperforms Geo-III, it is still far from Darwin. This is most probably due to \scavenger's current eagerness to restart after every conflict, whereas Darwin backtracks more carefully (cf. Sections \ref{sec:Implementation} and \ref{sec:Conclusion}). \scavenger and Darwin also treat variables in decision literals differently. Consequently, \scavenger detects more (and non-ground) conflicts, but learning conflict-driven clauses can be more expensive, because unifiers must be collected from the conflict graph and composed.

%For example, if $p(X)$ is chosen as a decision literal and later $\neg p(a)$ is propagated, \scavenger will detect the conflict, learn a conflict-driven clause and backtrack/restart. For Darwin, on the other hand, the $X$ in the decision literal $p(X)$ is not a usual universally quantified variable but is rather a \emph{parameter} with a \emph{default} semantics: the decision literal $p(X)$ means that $p(t)$ is assumed to be true for all instantiations $t$ of $X$ as long as $\neg p(t)$ is not in the model. Therefore, Darwin sees no conflict between $p(X)$ and $\neg p(a)$, and has to continue propagating until a \emph{ground} conflict is reached. Intuitively, \scavenger's ability to detect non-ground conflicts (and hence detect conflict more quickly) might give it an advantage when solving unsatisfiable problems. On the other, \scavenger's computation of conflict-driven learned clauses in the case of non-ground conflicts is more involved and computationally more expensive, since it requires a traversal of the conflict graph, collecting and composing unifiers. 
%Finally, it should be noted that Darwin's strategy (i.e. no backtracking and no clause learning in case of non-ground conflicts) is compatible with the Conflict Resolution calculus. In other words, the Conflict Resolution calculus does not enforce \scavenger's current proof search strategy, but allows it.

EP-\scavenger solved 28.2\% more problems than TD-\scavenger and 13.9\% more than PD-\scavenger. This suggests that non-termination of unit-propagation is an uncommon issue in practice: EP-\scavenger is still able to solve many problems, even though it does not care to bound propagation, whereas the other two variants solve fewer problems because of the overhead of bounding propagation even when it is not necessary. Nevertheless, there were 28 problems solved only by PD-\scavenger and 26 problems solved only by TD-\scavenger (among \scavenger's variants).
EP-\scavenger and PD-\scavenger can solve 9 problems with TPTP difficulty rating 0.5, all from the SYN and FLD domains. 3 of the 9 problems were solved in less than 10 seconds.

% \marginpar{Daniyar thinks this explanation is unlikely }
%A possible explanation for the poor performance is a suboptimal choice of initial depth threshold and strategy for increasing the threshold. This explanation seems to be supported by the stairs-like shape of the curve for PD-\scavenger. The shape suggests that there are about 80 problems that can be solved with propagation depth threshold equal to 0. Most of these problems can be solved quite quickly, in less than 12 seconds.  Then there are about 100 problems (180 minus 80) that can only be solved with propagation depth threshold equal to 1. But before increasing the threshold to 1, PD-\scavenger spends about 13 seconds trying useless propagations of depth 0, resulting in a steep step at $\#_{\mathrm{problems}} = 80$. Another step happens at $\#_{\mathrm{problems}} = 180$ (probably at the point when PD-\scavenger is wasting time doing propagations of depth 1 on problems that need propagations of depth greater than or equal to 2).

% TODO: Include plots for Satisfiable EPR CNF no-Equality benchmarks and 
% for all Unsatisfiable CNF no-equality benchmarks. In each of these figures include all provers that appear on Figure 2 above. Include EP-\scavenger on the plot for the non-EPR benchmarks as well. 

\section{Conclusions and Future Work}
\label{sec:Conclusion}

\scavenger is the first theorem prover based on the new Conflict Resolution calculus. The experiments show a promising, albeit not yet competitive, performance. 

A comparison of the performance of the three variants of \scavenger shows that it is non-trivial to interleave decisions within possibly non-terminating unit-propagations, and further research is needed to determine (possibly in a problem dependent way) optimal initial depth thresholds and threshold incrementation strategies. Alternatively, entirely different criteria could be explored for deciding to make an eager decision before propagation is over. For instance, decisions could be made if a fixed or dynamically adjusted amount of time elapses.

The performance bottleneck that needs to be most urgently addressed in future work is backtracking and restarting. Currently, all variants of \scavenger restart after every conflict, keeping derived conflict-driven clauses but throwing away the model construct so far. They must reconstruct models from scratch after every conflict. This requires a lot of repeated re-computation, and therefore a significant performance boost could be expected through a more sensible backtracking strategy. \scavenger's current naive unification algorithm could be improved with term indexing~\cite{Indexing}, and there might also be room to improve \scavenger's rough first-order analogue for the \emph{watched literals} data structure, even though the first-order challenges make it unlikely that something as good as the propositional watched literals data structure could ever be developed. Further experimentation is also needed to find optimal values for the parameters used in \scavenger for governing the initial thresholds and their incrementation policies. 
%Furthermore 
%\scavenger inherits from the proof compression system \texttt{Skeptik}~\cite{Skeptik} many data structures that had been implemented aiming at convenient proof manipulation instead of efficient theorem proving.

\scavenger's already acceptable performance despite the implementation improvement possibilities just discussed above indicates that automated theorem proving based on the Conflict Resolution calculus is feasible. However, much work remains to be done to determine whether this approach will eventually become competitive with today's fastest provers.

\paragraph{Acknowledgments: } We thank Ezequiel Postan for his implementation of TPTP parsers for \texttt{Skeptik}~\cite{Skeptik}, which we have reused in \scavenger. We are grateful to Albert A. V. Giegerich, Aaron Stump and Geoff Sutcliffe for all their help in setting up our experiments in StarExec. This research was partially funded by the Australian Government through the Australian Research Council and by the Google Summer of Code 2016 program. Daniyar Itegulov was financially supported by the Russian Scientific Foundation (grant 15-14-00066).

\bibliographystyle{splncs}
\bibliography{biblio}

\end{document}